\documentclass[twocolumn,showpacs,superscriptaddress,amsmath,amssymb]{revtex4-1}
\usepackage{graphicx}
\usepackage{dcolumn}
\usepackage{bm}
\usepackage{color}
\usepackage{epsfig}
\usepackage{epstopdf}
\usepackage{mathrsfs}
\usepackage{hyperref}
\usepackage{amsmath}
\usepackage{amsxtra}
\usepackage{amsfonts}

\usepackage{multirow}
\usepackage{makecell}
\usepackage{longtable}
\usepackage{diagbox}
\usepackage{slashbox}

\def\bd{
\begin{document}} \def\ed{\end{document}}
\def\bmp{\begin{minipage}} \def\emp{\end{minipage}}
\def\bcc{\begin{center}} \def\ecc{\end{center}}     \def\npg{\newpage}
\def\beq{\begin{equation}} \def\eeq{\end{equation}} \def\hph{\hphantom}
\def\be{\begin{equation}} \def\ee{\end{equation}} \def\r#1{$^{[#1]}$}
\def\n{\noindent} \def\ni{\noindent} \def\pa{\parindent}
\def\hs{\hskip} \def\vs{\vskip} \def\hf{\hfill} \def\ej{\vfill\eject}
\def\cl{\centerline} \def\ob{\obeylines}  \def\ls{\leftskip}
\def\underbar#1{$\setbox0=\hbox{#1} \dp0=1.5pt \mathsurround=0pt
   \underline{\box0}$}   \def\ub{\underbar}    \def\ul{\underline}
\def\f{\left} \def\g{\right} \def\e{{\rm e}} \def\o{\over} \def\d{{\rm d}}
\def\vf{\varphi} \def\pl{\partial} \def\cov{{\rm cov}} \def\ch{{\rm ch}}
\def\la{\langle} \def\ra{\rangle} \def\EE{e$^+$e$^-$} \def\pt{p_{\rm t}}
\def\pti{p_{{\rm t},i}} \def\vti{v_{{\rm t},i}}
\def\ptj{p_{{\rm t},j}}\def\Pt{P_{\rm t}} \def\vt{v_{\rm t}}

\def\bitz{\begin{itemize}} \def\eitz{\end{itemize}}
\def\btbl{\begin{tabular}} \def\etbl{\end{tabular}}
\def\btbb{\begin{tabbing}} \def\etbb{\end{tabbing}}
\def\beqar{\begin{eqnarray}} \def\eeqar{\end{eqnarray}}
\def\\{\hfill\break} \def\dit{\item{-}} \def\i{\item}
\def\bbb{\begin{thebibliography}{9} \itemsep=-1mm}
\def\ebb{\end{thebibliography}} \def\bb{\bibitem}
\def\bpic{\begin{picture}(260,240)} \def\epic{\end{picture}}
\def\akgt{\cl{\bf ACKNOWLEDGMENTS}}
\def\fgn{\noindent{\bf\large\bf figure captions}}
\def\m1{\langle N_p\rangle} \def\u2{\langle N_{\bar p}\rangle} \def\Nap{N_{\bar
p}}
\def\lan{\langle}
\def\ran{\rangle}
\def\p{\pi}
\def\ifmath#1{\relax\ifmmode #1\else $#1$\fi}%
\def\rc{\ifmath{{\mathrm{c}}}}
\def\cut{\ifmath{{\mathrm{cut}}}}
\def\rF{\ifmath{{\mathrm{F}}}}
\def\rK{\ifmath{{\mathrm{K}}}}
\def\rp{\ifmath{{\mathrm{p}}}}
\def\rt{\ifmath{{\mathrm{t}}}}
\def\LAB{\ifmath{{\mathrm{LAB}}}}
\def\cut{\ifmath{{\mathrm{cut}}}}
\def\beq{\begin{equation}}
\def\eeq{\end{equation}}

\newcommand{\cinst}[2]{$^{\mathrm{#1}}$~#2\par}
\newcommand{\crefi}[1]{$^{\mathrm{#1}}$}
\newcommand{\crefii}[2]{$^{\mathrm{#1,#2}}$}
\newcommand{\crefiii}[3]{$^{\mathrm{#1,#2,#3}}$}
\newcommand{\HRule}{\rule{0.5\linewidth}{0.5mm}}
\newcommand{\minitab}[2][l]{\begin{tabular}{#1}#2\end{tabular}}

\bd
\title{The method of mixed events for higher cumulants of conserved charges}

\author{Fan Zhang}
\affiliation{Key Laboratory of Quark and Lepton Physics (MOE) and
Institute of Particle Physics, Central China Normal University, Wuhan 430079, China}
\author{Zhiming Li}
\affiliation{Key Laboratory of Quark and Lepton Physics (MOE) and
Institute of Particle Physics, Central China Normal University, Wuhan 430079, China}
\author{Lizhu Chen}
\affiliation{School of Physics and Optoelectronic Engineering, Nanjing University of Information Science and Technology, Nanjing 210044, China}
\author{Xue Pan}
\affiliation{School of Electronic Engineering, Chengdu Technological University, Chengdu 611730, China}
\author{Mingmei Xu}
\affiliation{Key Laboratory of Quark and Lepton Physics (MOE) and
Institute of Particle Physics, Central China Normal University, Wuhan 430079, China}
\author{Yeyin Zhao}
\affiliation{Key Laboratory of Quark and Lepton Physics (MOE) and
Institute of Particle Physics, Central China Normal University, Wuhan 430079, China}
\author{Yu Zhou}
\affiliation{ Department of Physics and Astronomy University of California Los Angeles, CA 90095, USA}
\author{Yuanfang Wu}\email{wuyf@mail.ccnu.edu.cn}
\affiliation{Key Laboratory of Quark and Lepton Physics (MOE) and
Institute of Particle Physics, Central China Normal University, Wuhan 430079, China}

\begin{abstract}
Higher cumulants of conserved charges are sensitive observables of quantum chromodynamics phase transitions. The sample of mixed events provides a background to estimate non-critical effects of cumulants. Four possible methods for constructing the sample of mixed events are suggested. The effectiveness of each method is examined. It is showed that the method of most random or least constrain is the best, rather than the conventional method.
\end{abstract}


\maketitle

\section{Introduction}

Higher cumulants of conserved charges are suggested sensitive observables of quantum chromodynamics (QCD) phase transitions~\cite{m7,m8,m9,m10,m11,m12,m13}. They draw much of our attention in relativistic heavy-ion collisions (RHIC)~\cite{star-prl1, star-prl2}. Their non-monotonic dependence of incident energy has been observed at the RHIC beam energy scan (BES I) and has been considered as a potential signal of the critical point (CP)~\cite{scan-a,scan-b,scan-c,scan-d,scan-e}. 

As we know, there are non-critical fluctuations in higher cumulants. Numerous efforts have been made in subtracting non-critical effects~\cite{Lizhu-JPG1,Lizhu-JPG2, Panx-PRC, Zhoudm-PRC,Koch-PRC91-027901, acceptenc, Luoxf-CBWC, Songhc-PRC,Songhc-PRC2}.  

Critical fluctuations come from inner correlations between particles of an event, where the correlation length is divergent. Non-critical fluctuations have two kinds of sources. One is caused by conventional mechanisms, such as resonance decay, collective flow, global conservation of energy, momentum, and charges and so on. The length of these conventional correlations is finite and fixed. Their fluctuations are usually small in comparison to critical fluctuations, and result in a constant shift in cumulants~\cite{Songhc-PRC}.

Another source of non-critical fluctuations is global and systematic effects, such as statistical fluctuations due to insufficient number of particles~\cite{Panx-PRC, Zhoudm-PRC}, initial size fluctuations for different impact parameters~\cite{Songhc-PRC}, centrality bin width~\cite{Luoxf-CBWC}, detection efficiency and experimental acceptance cuts~\cite{Koch-PRC91-027901, acceptenc}. These global and systematic effects are independent of inner correlations between particles of an event.

In order to deduct one of these global and systematic effects, usually a specified scheme of correction is suggested. For example, statistical fluctuations are estimated by Poisson distribution~\cite{Lizhu-JPG2, Panx-PRC}. To deduct the influence of centrality bin width, a well known scheme, centrality bin width correction (CBWC), is proposed~\cite{Luoxf-CBWC}. In order to subtract the influence of detection efficiency, a complex formula which connects the true cumulant to the actually measured cumulant is introduced~\cite{acceptenc}. 

For a real data sample, all those global and systematic effects are involved. It is difficult to subtract one of them. A good scheme of subtraction should take all of them together into account. The {\it sample of mixed events} just provides a background of such~\cite{NA491,NA492, Zhoudm-PRC}. 

Conventionally, the particles of a mixed event are randomly selected from different events~\cite{NA491,NA492,Zhoudm-PRC}. Meanwhile, it requires that the total multiplicity distribution of mixed events keeps consistent with that of the original sample. So that the global and systematic feature of the original sample retains in the sample of mixed events. 

In relativistic heavy ion collisions, the idea of mixed events has been applied to various measurements, such as, two-particles correlations~\cite{STAR-2y}, the ratio of particle production~\cite{STAR-ratio1}, transverse momentum spectrum~\cite{STAR-pt}, elliptic flow~\cite{STAR-flow1,STAR-flow2}, and so on. The method of mixed events changes with observable. 

For example, two-particle rapidity correlations, what concern is rapidity positions of particles of an event. The method of mixed events is to replace true rapidity positions of all particles of an event by those chosen  randomly from different events~\cite{STAR-2y}. While, for cumulants of conserved charges, what concern is the number of charged particles of an event. So the task of mixed events is to turn off inner correlations between charged particles of an event, and the correlation of charged particles with its associated event. 
To full fill these requirements, there are numbers of methods in constructing mixed event. 

In this paper, we provide four possible methods in section II. Then, in section III, we exam the effectiveness of each method. It is showed that the most random or least constrain method is the best, rather than the conventional method. A brief summary is presented in section IV.

\section{Four possible methods}

Higher cumulants of conserved charges are defined as variance ($\sigma^2$), skewness ($S$), kurtosis ($\kappa$), and their products, $S\sigma$ and $\kappa\sigma^2$, i.e.,
\begin{equation}\label{cumulants}
\begin{split}
\sigma^2 &=\langle(\triangle N_{\rm c})^2\rangle, \\
S &=\langle(\triangle N_{\rm c})^3\rangle/\sigma^3,\\
\kappa &=\langle(\triangle N_{\rm c})^4\rangle/\sigma^4-3,\\
S\sigma &=\langle(\triangle N_{\rm c})^3\rangle/\sigma^2,\\
\kappa\sigma^2 &=\langle(\triangle N_{\rm c})^4\rangle/\sigma^2-3\sigma^2.
\end{split}
\end{equation} 
Where the average $\langle\rangle$ is over the whole event sample. $\triangle N_{\rm c}=N_{\rm c}-\langle N_{\rm c}\rangle$. $N_{\rm c}$ is the number of particles with conserved charges. In general, conserved charge refers to baryon, strangeness, or electric charge. In this paper, we restrict the conserved charge to baryon, or strangeness only. The total number of particles of an event is electric charged, i.e, $N_{\rm ch}$,  multiplicity. 

In Eq.~(\eqref{cumulants}), $N_{\rm c}$ is associated with event.  For a given event,
$N_{\rm c}$ charged particles correlate with each other and with other non-charged $N_{\rm ch}-N_{\rm c}$ particles. The distribution of $N_{\rm c}$ well present this kind of correlation. 
Higher cumulants of charged particles measure all these correlations. For a mixed event, all these correlations has to be removed. Meanwhile, global characters of mixed events should retain, i.e., the distribution of $N_{\rm ch}$ , and the $\langle N_{\rm c} \rangle$ should keep consistent with those of the original sample. 

To keep the multiplicity distribution, we simply take event multiplicity $N_{\rm ch}$ from the original sample. Conventionally, each particle of a mixed event is randomly taken from $N_{\rm ch}$ different events~\cite{NA491,NA492, Zhoudm-PRC}. There are two possible ways to take $N_{\rm ch}$ particles from different events.  For the Method-I, we can select randomly $N_{\rm ch}$ events from the original sample, and take one particle from each selected event. 

In this case, each particle of a mixed event comes from different events and has no correlation with others. The distribution of charged particles $N_{\rm c}$ of mixed sample is,
\begin{equation}\label{Nc-dis, method-I}
    P^{\rm m}(N_{\rm c})=\sum\limits_{N_{\rm ch}=N_{\rm c}}^\infty P^{\rm o}(N_{\rm ch})\left(
                                                            \begin{array}{c}
                                                              N_{\rm ch}\\
                                                              N_{\rm c}\\
                                                            \end{array}
                                                          \right)\prod\limits_{i=1}^{N_{\rm c}}p^i_{\rm c} \prod\limits_{j=1}^{N_{\rm ch}-N_{\rm c}} p^j_{\rm nc}  
\end{equation}
\noindent Where, the probability of getting a charged particle in the $i$th original event is,
\begin{equation}\label{pc, method-I}
p^i_{\rm c}=\frac{N_{\rm c}^i}{N_{\rm ch}^i}P^o(N_{\rm c}^i)P^o(N_{\rm ch}^i),
\end{equation}
\noindent and the probability of getting a non-charged particle in the $j$th original event is,
\begin{equation}\label{pnc, method I}
p^j_{\rm nc} =(1-\frac{N_{\rm c}^j}{N_{\rm ch}^j})P^o(N_{\rm c}^j)P^o(N_{\rm ch}^j)
\end{equation}
Here, for the $i$th event, $\frac{N_{\rm c}^i}{N_{\rm ch}^i}$ is a constant. $P^o(N_{\rm c})$ and $P^o(N_{\rm ch})$ are distributions of original number of charged particles and multiplicity, respectively. If the statistics of the original sample is large enough, the mean of $N_{\rm c}$ of mixed sample should be approximately consistent with that of the original sample.

In this Method-I, one particle in the original sample may be used twice, or more. In order to avoid repeating, the used particle can be taken away from the original sample. In the case (labelled as
Method-II), the probability of getting a charged particle from the $i$th original event, i.e.,  Eq.~(\ref{Nc-dis, method-I}) becomes,
\begin{equation}\label{pc, method-II}
p^i_{\rm c}=\frac{N^{i'}_{\rm c}}{N^{i'}_{\rm ch}}P^{\rm o}(N^{i'}_{\rm c})P^{\rm o}(N^{i'}_{\rm ch}),
\end{equation}
\noindent and the probability of getting a non-charged particle from the $j$th original event is,
\begin{equation}\label{pnc, method-II}
p^j_{\rm nc} =(1-\frac{N^{j'}_{\rm c}}{N^{j'}_{\rm ch}})P^o(N^{j'}_{\rm c})P^o(N^{j'}_{\rm ch})
\end{equation}
Here, for the $i$th given event, $\frac{N^{i'}_{\rm c}}{N{i'}_{\rm ch}}$ is no longer a constant. $N^{i'}_{\rm c}$ and $N^{i'}_{\rm ch}$ are left numbers of particles with conserved charges and electric charges, respectively. They both decrease with used number of particles. For the Method-II, the average number of $N_{\rm c}$  of the mixed sample is exactly identical to that of the original sample.

In fact, if the original sample is large enough, the number of particles from all events can be considered to be infinitely large. We can put all of them into a pool. Then randomly take $N_{\rm ch}$ particles from the pool. These $N_{\rm ch}$ particles are approximated from different events and have no correlations with each other. In the case (labelled as Method-III), the distribution of charged particles of mixed sample is,
\begin{equation}\label{Nc-dis, pool I}
    {P^{\rm m}(N_c)=\sum\limits_{N_{\rm ch}=N_{\rm c}}^\infty P^{\rm o}(N_{ch})\left(
                                                            \begin{array}{c}
                                                              N_{ch}\\
                                                              N_c\\
                                                            \end{array}
                                                          \right)p_c^{N_c} p_{nc}^{N_{ch}-N_c}
   }
\end{equation}
Where,
\begin{equation}\label{pc, pool I}
p_c=\frac{\sum\limits_{i=1}^{N_{event}} N_{c}^{i}}{\sum\limits_{i=1}^{N_{event}}N_{ch}^{i}},
\end{equation}
and so is
\begin{equation}\label{pnc, pool I}
p_{nc} =1-\frac{\sum\limits_{i=1}^{N_{event}} N_{c}^{i}}{\sum\limits_{i=1}^{N_{event}}N_{ch}^{i}}.
\end{equation}
Where the probability of taking a charged particle from the pool is a constant, and independent of distributions of original charged particles and multiplicity. If the statistics of the original sample is large enough, $\langle N_{\rm c}\rangle$ of the mixed sample should be approximately consistent with that of the original sample.

If we require that each particle in the pool appears once in the mixed events, then used particles should not be put back into the pool again. In the case (labelled as Method-IV), the probability of charged particles of mixed sample, i.e., Eq.~(\ref{pc, pool I}) or Eq.~(\ref{pnc, pool I}) will change with used particles, and $\langle N_{\rm c}\rangle$ of mixed sample should be exactly identical with that of the original sample.  

Obviously, Methods-II and Method-VI are more restrictive in comparison to Methods-I and Method-III. Whether such a restriction is necessary at current statistics is not clear and should be examined. Method-III is more random or the least constrain and simple than the conventional method-I, although it is not obvious whether the Method-III is effective. So we will examine each of them in the following.

\section{The effectiveness of four methods}

Multiplicity of mixed sample for above four methods is directly taken from the original sample. So the multiplicity distribution is well preserved. The remaining concerns are $\langle N_{\rm c}\rangle $ of mixed sample, the correlation of $N_{\rm c}$ and its associated mixed event, the influence of the original distribution of $N_{\rm c}$, and the statistics required for each of methods. We will first examine the $\langle N_{\rm c}\rangle $ of four mixed samples at current statistics.

\subsection{ $\langle N_{\rm c}\rangle $ of four mixed samples}

As described in the above section, total numbers of $N_{\rm c}$ of mixed Sample-II and Sample-IV are exactly identical with that of the original sample. The approximation only appears in mixed Sample-I and Sample-III, and depends on the total number of events and multiplicity. To see $\langle N_{\rm c}\rangle $ in four mixed samples, we make a simple model simulation, where multiplicity and charged particles are both produced by a Poisson distribution. The mean of multiplicity is 100, which is in the range of the RHIC BES plan. The mixed sample is constructed respectively by four methods mentioned above. The total number of events is one million, the lowest statistics at the RHIC BES I.  

$\langle N_{\rm c}\rangle$ of the original sample, and four mixed samples are presented in Table I. $\langle N_{\rm c}\rangle$ of the original sample is 20. $\langle N_{\rm c}\rangle$ for mixed sample-II and sample-IV are expected, exactly identical with that of the original sample. Within statistical error, $\langle N_{\rm c}\rangle$  of mixed sample-I and sample-III are also identical with that of the original sample. So for the statistics at the RHIC BES I or II,  $\langle N_{\rm c}\rangle$ of mixed samples given by four mixed methods are all consistent with that of the original sample.

\begin{table}[h]
  \caption{$\langle N_c\rangle$ of the original and mixed samples.}
\begin{tabular}{|c|c|c|c|c|c|}

  \hline
 \multirow{2}{*}{Sample} & \multirow{2}{*}{Original} & \multirow{2}{*}{Mixed-I} & \multirow{2}{*}{Mixed-II} & \multirow{2}{*}{Mixed-III} &  \multirow{2}{*}{Mixed-IV} \\

     \multirow{2}{*}{}& \multirow{2}{*}{} & \multirow{2}{*}{} &  \multirow{2}{*}{} & \multirow{2}{*}{}& \multirow{2}{*}{}\\
    \hline
\multirow{2}{*}{$\langle N_c\rangle$} &  \multirow{2}{*}{20} &  \multirow{2}{*}{$20.009$} &  \multirow{2}{*}{20} &  \multirow{2}{*}{$19.98$} &  \multirow{2}{*}{20} \\

\multirow{2}{*}{} &  \multirow{2}{*}{} &  \multirow{2}{*}{$\pm 0.007$} &  \multirow{2}{*}{} &  \multirow{2}{*}{$\pm 0.006$} &  \multirow{2}{*}{} \\
     \multirow{2}{*}{}& \multirow{2}{*}{} & \multirow{2}{*}{} &  \multirow{2}{*}{} & \multirow{2}{*}{}& \multirow{2}{*}{}\\

  \hline
\end{tabular}
\end{table}

\subsection{Concerning correlation in four mixed samples}

In order to check the correlation between $N_{\rm c}$ and its associated event, we designed a toy model. In the model, the original event is labelled by a sequence number $N^e$. $N^e$ is related linearly to the number of charged particles $N^e_{\rm c}$, i.e., $N^e_{\rm c}=aN^e$ and proportional constant $a=0.004$. The multiplicity is fixed to $N_{\rm ch}=400$, so that $N^e_{\rm c}\le N_{\rm ch}$. The total number of events $N_{\rm event}$ is 0.1 million. 

Then, using the above mentioned four methods, we can produce corresponding mixed samples. For each mixed sample, we calculate the correlation coefficient of $N^e_{\rm c}$ and $N^e$, i.e.,
\begin{equation}\label{c-coefficient}
  C(N^e,N^e_{\rm c},)=\frac{\langle N^e\cdot N^e_{\rm c}\rangle}{\langle N^e\rangle \langle N^e_{\rm c} \rangle }-1.
\end{equation}
Where the average is over all events. if $N^e_{\rm c}$ is independent of $N^e$, $C(N^e, N^e_c,) =0$.

$C(N^e,N^e_{\rm c}) $ of the original and four mixed samples are presented in Table II. It is 0.998 in
the original sample. This shows that $N^e$ linearly correlates to $N^e_{\rm c}$, as provided in the model. For four mixed samples, they are all close to zero within error. So all four methods successfully eliminate the original correlations between charged particles and its associated event.

   \begin{table}[h]
   \centering
   \caption{$ C(N^e, N^e_c)$ of the original and four mixed samples.}
\begin{tabular}{|c|c|c|c|c|c|}

  \hline
 \multirow{2}{*}{Sample} & \multirow{2}{*}{Original} & \multirow{2}{*}{Mixed-I} & \multirow{2}{*}{Mixed-II} & \multirow{2}{*}{Mixed-III} &  \multirow{2}{*}{Mixed-IV} \\
     \multirow{2}{*}{}& \multirow{2}{*}{} & \multirow{2}{*}{} &  \multirow{2}{*}{} & \multirow{2}{*}{} & \multirow{2}{*}{} \\
    \hline
\multirow{2}{*}{$C(N^e, N^e_c)$} &  \multirow{2}{*}{$0.998$} &  \multirow{2}{*}{$0.007$} &  \multirow{2}{*}{$0.009$} &  \multirow{2}{*}{$0.004$} &  \multirow{2}{*}{$0.011$} \\
\multirow{2}{*}{} &  \multirow{2}{*}{} &  \multirow{2}{*}{$\pm 0.01$} &  \multirow{2}{*}{$\pm 0.014$} &  \multirow{2}{*}{$\pm 0.006$} &  \multirow{2}{*}{$\pm 0.015$} \\

     \multirow{2}{*}{}& \multirow{2}{*}{} & \multirow{2}{*}{} &  \multirow{2}{*}{} & \multirow{2}{*}{} & \multirow{2}{*}{} \\

  \hline
\end{tabular}
\end{table}

\subsection{Influence of the original distribution of $N_{\rm c}$}

Among higher cumulants defined in Eq.~(\ref{cumulants}), $\kappa\sigma^2$ is most sensitive to the distribution of $N_{\rm c}$. In order to check if the original distribution of $N_{\rm c}$ influences $\kappa\sigma^2$ of the mixed sample, we study the $\kappa\sigma^2$ of the mixed sample for two very different original distributions of $N_{\rm c}$ at various statistics. 

Two distributions of $N_{\rm c}$ are Poisson and summation of two $\delta$ functions. Where the mean of charged particles for Poisson distribution is 30. For two $\delta$ functions, we set $N_{\rm c1}=15$ and $N_{\rm c2}=45$, respectively, i.e., $P(N_{\rm c})=\frac{1}{2}\delta(15)+\frac{1}{2}\delta(45)$. This implies that $N_{\rm c}$ can only be two possible values, i.e., either 15, or 45. It is very specified and different from Poisson distribution.

Distributions of original multiplicity for these two cases are still $\delta(100)$ distribution,this implies that $N_{\rm ch}$ can only be 100. Using four methods mentioned above, mixed samples for two cases are constructed. 

Statistics dependences of $\langle \kappa\sigma^2\rangle $ of mixed samples given by Method-I and Method-III are presented in Fig.~1(a) and (b). Where open black squares and solid red points are presented for   Poisson distribution and two-$\delta$ functions as the original distributions of $N_{\rm c}$. The blue horizontal line is the expectation of analytic calculation.
 
In order to keep the consistent of statistical errors for different statistics, we let the total number of events be $10^6$ and the same for all points. Where the mean is over a given statistics in horizontal coordinates, and over the sub-samples with the same statistics, as suggested in ref. ~\cite{Lizhu-JPG1,Lizhu-JPG2}.
  
\begin{figure}[htb]
\includegraphics[width=0.5\textwidth]{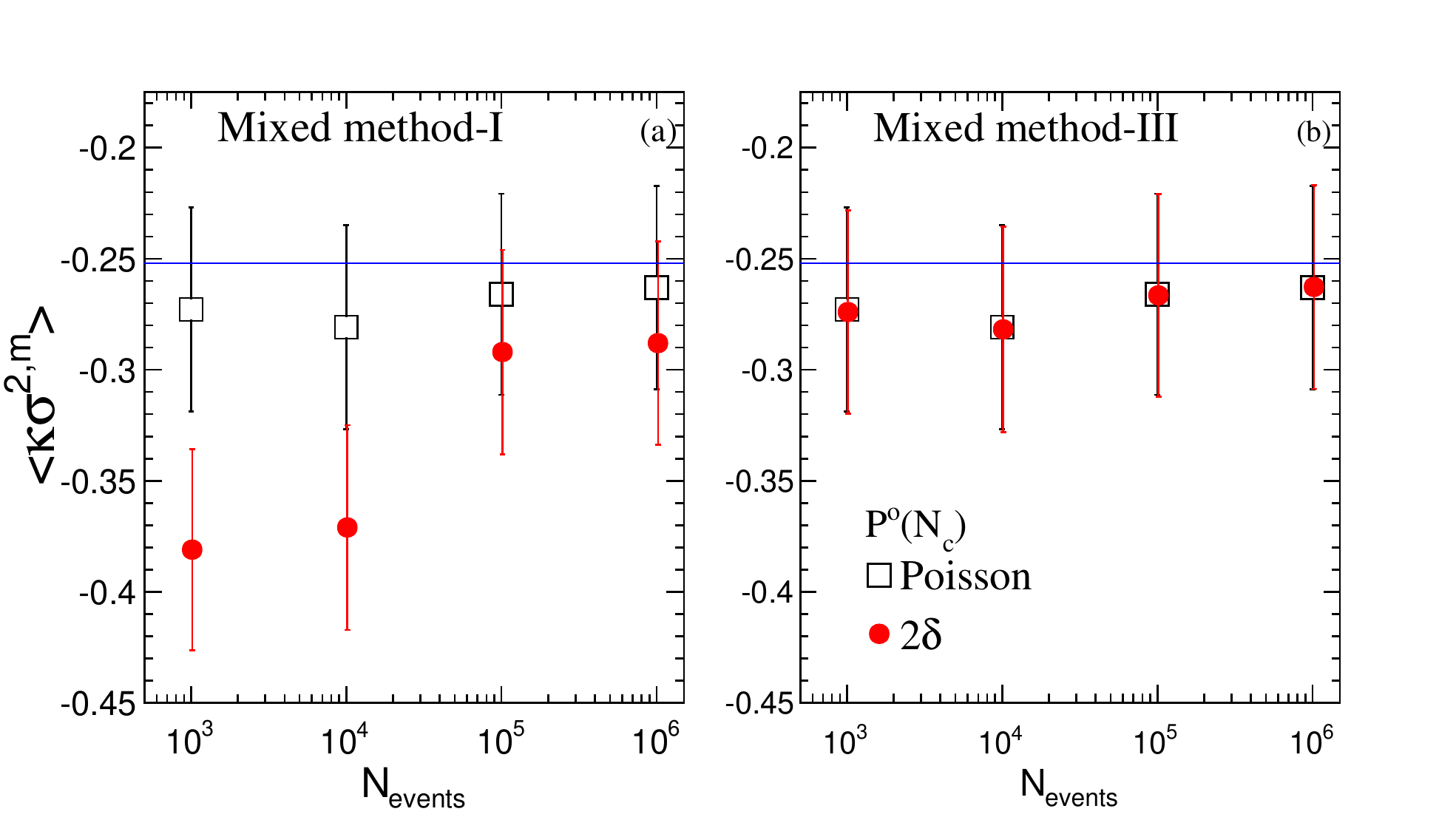}
\caption{\label{Fig. 1}(Color online). Statistic dependences of $\kappa\sigma^2$ of mixed samples-I (a) and Sample-III (b), where the original distribution of $N_{\rm c}$ are Poisson (open black squares) and summation of two-$\delta$ functions (solid red points). } 
\end{figure}

For the mixed Method-I, at low statistics, solid red points in Fig.~1(a) deviate from open black squares, and are much smaller than expectation, i.e., the blue line. When statistics are larger than $10^5$, for a given statistics, solid red point and open black square overlap within error, and approach the blue line within error. So when statistics are not large enough, the original distribution of $N_{\rm c}$ still influences the $\kappa\sigma^2$ of the mixed sample. This is understandable. As showed in Eq~(\ref{Nc-dis, method-I}), Eq~(\ref{pc, method-I}), and Eq~(\ref{pnc, method I}), the probability of getting a particle with conserved charge from the $i$th original event is determined by the original distribution of $N_{\rm c}$. So via each particle which is selected from an original event, the original distribution of $N_{\rm c}^i$ , i.e., $P^o(N_{\rm c})$, influences the distribution of $N_{\rm c}$ of the mixed sample.

While, for the mixed Method-III, solid red point well overlap with black open square at every statistics, as showed in Fig.~1(b). So even at the lowest statistics, $10^3$, $\kappa\sigma^2$ of the mixed sample is independent of the original distribution of $N_{\rm c}$. In this method, the probability of getting a particle with conserved charge from the pool is independent of the original distribution of $N_{\rm c}$, as showed in Eq.~(\ref{Nc-dis, pool I}), Eq.~(\ref{pc, pool I}), and Eq.~(\ref{pnc, pool I}), where $P^o(N_{\rm c})$ has no chance to appear. The information of the distribution of original $N_{\rm c}$ is completely lost in a mixed event. 

The statistical dependences of $\langle \kappa\sigma^2\rangle $ of mixed sample-II and sample-IV are not presented, and similar to Fig.~1(a) and (b), respectively. This shows again that the Method-II and Method-IV are unnecessary. It is trivial to require each particle in the original sample to appear only once in a mixed sample. The randomness, or forgetting concerning correlations of the original event, is essential in constructing a mixed event. Therefore, the Method-III is the best among four methods. 

\section{Summary}

In this paper, we study the effectiveness of four possible methods for constructing the sample of mixed events. Four methods are all equally good in removing the correlation between the number of charged particles and its associated event, and preserving the global and systematic characters of the original sample, such as the mean of charged particles, and multiplicity distribution. Differences among four methods are
in eliminating the influence of the original distribution of $N_{\rm c}$.

For conventional Method-I, i.e., each particle of mixed event is randomly taken from different original events, it shows that cumulants of the mixed sample is still influenced by the original distribution of $N_{\rm c}$ at low statistics. The influence becomes negligible only at high statistics.

For the Method-III, all particles of the original events are put into a pool, and each particle of the mixed event is randomly taken from the pool. Obviously, this method is the least constrain and more random. It shows in this case that cumulants of the mixed sample is free from the influence of the original distribution of $N_{\rm c}$ at even very low statistics.  

For the Methods II and Method-IV, an additional requirement is that original particles appear only once in the mixed sample. This requirement is trivial and unnecessary. Therefore, Method-III is the most random or least constrain, and is the best, rather than the conventional Method-I, or Methods-II and Method-IV. 

The mixed sample provides a good estimation for global and systematic non-critical effects in higher cumulants of conserved charges, such as, the statistical fluctuations, initial size fluctuations, centrality bin width corrections (CBWC), detection efficiency, and experimental acceptance cuts. Some of them are demonstrated in a coming paper ref.~\cite{Zhangf-mix-2}.

\section{Acknowledgement}

This work is supported in part by the Ministry of Science and Technology (MoST) under grant No. 2016YFE0104800, and the Fundamental Research Funds for the Central Universities under
grant No. CCNU19ZN019.

\bibliographystyle{unsrt}
\bibliography{method}

\ed